\documentclass[pre,twocolumn,showpacs]{revtex4}
\usepackage[pctex32]{graphics}
\begin{document}
\title{Consistency in the driven butterfly}

\author{Atsushi Uchida$^{1}$}
\author{Kazuyuki Yoshimura$^{2}$}
\author{Peter Davis$^{2}$}
\author{Shigeru Yoshimori$^{1}$}
\author{Rajarshi Roy$^{3}$}
\affiliation{$^{1}$Department of Electronics and Computer Systems, Takushoku University, 815-1 Tatemachi, Hachioji, Tokyo 193-0985, Japan}
\affiliation{$^{2}$NTT Communcation Science Laboratory, NTT Corporation, 2-4 Hikaridai, Seika-cho, Soraku-gun, Kyoto, 619-0237 Japan}
\affiliation{$^{3}$IPST, University of Maryland, College Park, Maryland 20742 USA}

\date{\today}

\begin{abstract} 
We study the consistency of the responses in the chaotic Lorenz model driven by chaotic and colored-noise signals. A measure is introduced for the consistency of the response in the system for repeated applications of the same drive signal, starting at different initial conditions of the response system. The convergence of responses is analyzed in dynamical phase space in terms of the local conditional Lyapunov exponent. The dependence of the local conditional Lyapunov exponent on drive signal, and the existence of multiple basins of consistency depending on strength of drive signal, are shown.
\end{abstract}

\maketitle

The dynamical response of complex structures ranging from biomolecules to the Millennium Bridge~\cite{strogatz43} to complex drive waveforms is an important issue for science and engineering. Analysis of dynamical responses can provide a means for non-invasive testing and evaluation of the integrity of structures and changes due to damage or aging~\cite{moniz036215}. This approach may be extended to characterize the behavior of systems with complex internal dynamics, such as electronic circuits, laser systems, musical instruments and even brains. However, characterizing the responses of nonlinear dynamical systems with complex dynamics presents many challenges. It may appear at first sight that the wide variety of possible responses for a system with complex dynamics would make this task hopeless. On the other hand, a system as complex as the brain can have responses with various degrees of reproducibility and consistency when subjected to repeated inputs of information and stimulation. It is our goal here to develop a new and practical approach to analyzing complex drive-response behavior based on measuring consistency~\cite{uchida244102} of responses to repeated applications of the same drive signal. 

Pecora and Carroll~\cite{pecora2374} analyzed the stability of the response of a dynamical system to a drive signal through evaluation of the conditional Lyapunov exponent, conditioned on the drive signal. In their examples, the drive waveforms were generated by low dimensional chaotic system, and the drive and response systems displayed identical synchronization~\cite{pecora821}. The more general case of stable synchronized response where drive and response waveforms are not identical has been referred to as generalized synchronization~\cite{rulkov980}. The condition of generalized synchronization corresponds to the condition for identical synchronization of two replica auxiliary systems driven by the same input signal. In the past decade, many studies have extended the drive signals to include noisy waveforms~\cite{maritan1451,zhou230602}, which are very high dimensional. Recent research on coherence resonance has included experiments on colored-noise driven excitable systems~\cite{coherence}. We also note that the concept of reliability has been used to describe the timing consistency of the evoked spike-sequences of rat cortical neurons in response to repeated presentations of the same noise waveform~\cite{mainen1503}. 


In this Letter, we consider the consistency viewpoint which emphasizes the reproducibility of responses to a particular input in driven nonlinear systems. We use the specific example of the Lorenz system~\cite{lorenz130} driven by chaos and noise signals. We show the onset of consistency in terms of a consistency measure and the conditional Lyapunov exponent~\cite{pecora2374}, which measures the convergence of trajectories in the space of responses conditioned on a particular drive signal. We also examine the local conditional Lyapunov exponent (LCLE)~\cite{witt5050} which allows us to see the contraction and expansion regions, and see the dependence of the distribution of the LCLE on changes of the drive signal. Finally, we consider the existence of multiple basins of consistent response for the same drive signal and the dependence of the basins on the drive signal.

Consider a dynamical system given by $d{\mbox{\boldmath X}}(t)/dt = {\mbox{\boldmath F}}({\mbox{\boldmath X}}(t), {\mbox{\boldmath S}}(t))$, where ${\mbox{\boldmath X}}$ is the state variables of the response system, ${\mbox{\boldmath S}}$ is the drive signal, and ${\mbox{\boldmath F}}$ is a nonlinear function. We consider one of the observable state variables $x(t)$. The consistency $C$ can be measured as the average correlation between responses for a set of response system initial conditions,
\begin{eqnarray}
\label{crosscorrelation}
C&=&\frac{2}{N(N-1)}\sum_{i=1}^{N}\sum_{j=1 (j > i)}^{N}\nonumber\\
& &\left[\frac{<(x(t;x_{i}(0))-\bar{x}_{i})(x(t;x_{j}(0))-\bar{x}_{j})>}{\sigma_{i}\sigma_{j}}\right],
\end{eqnarray}
where $x(t;{\mbox{\boldmath X}}_{i}(0))$ and $x(t;{\mbox{\boldmath X}}_{j}(0))$ are variables $x(t)$ starting from different initial conditions of ${\mbox{\boldmath X}}_{i}(0)$ and ${\mbox{\boldmath X}}_{j}(0)$, $\bar{x}_{i}$ and $\bar{x}_{j}$ are the mean values of the two time series of $x(t;{\mbox{\boldmath X}}_{i}(0))$ and $x(t;{\mbox{\boldmath X}}_{j}(0))$, and $\sigma_{i}$ and $\sigma_{j}$ are the standard deviations of the two time series. The angle brackets denote time averaging.



We use the Lorenz model~\cite{lorenz130} with an additive drive for the $y$-variable, $dy/dt=-xz+rx-y+Ds(t)$, where the drive signal $s(t)$ is multiplied by the coefficient of drive strength $D$. Two examples of $s(t)$ are (i) $y$ variable of the Lorenz chaos generated from the basic model and (ii) an exponentially correlated colored noise waveform generated using an Ornstein-Uhlenbeck process~\cite{fox5938} with a correlation time $\tau_{c}=0.2$. The temporal waveforms of a drive signal and corresponding response of the $x$ variable starting from ten different initial conditions are shown in Figs. 1(a) and 1(b) for the chaos and colored-noise drive signals. It can be seen that the waveforms of the responses converge after a short transient. We calculated the consistency measure $C$ of Eq.~(\ref{crosscorrelation}) for an ensemble of initial states randomly selected on the unperturbed Lorenz butterfly. Figures 1(c) and 1(d) show the asymptotic value of $C$ ($\rightarrow 1$) as a function of the drive strength $D$ for the Lorenz chaos and colored-noise drive signals.

To calculate the conditional Lyapunov exponent we begin with the linearized equations of $d{\mbox{\boldmath $\xi$}}(t)/dt = {\mbox{\boldmath D}}_{\tiny \mbox{\boldmath X}}{\mbox{\boldmath F}}({\mbox{\boldmath $\xi$}}(t), {\mbox{\boldmath X}}(t))$, where ${\mbox{\boldmath $\xi$}}$ are the linearized variables of ${\mbox{\boldmath X}}$ and ${\mbox{\boldmath D}}_{\tiny \mbox{\boldmath X}}{\mbox{\boldmath F}}$ is the Jacobian matrix with respect to ${\mbox{\boldmath X}}$ (not including ${\mbox{\boldmath S}}$). The conditional Lyapunov exponent $\lambda_{c}$ for consistency is defined as,
\begin{eqnarray}
\label{equation4}
\lambda_{c}=\lim_{T\rightarrow\infty}\frac{1}{T}\log\frac{\|{\mbox{\boldmath $\xi$}}(T)\|}{\|{\mbox{\boldmath $\xi$}}(0)\|},
\end{eqnarray}
where $\|{\mbox{\boldmath $\xi$}}\|$ is the norm of the linearized variables. We only consider the maximal value of the conditional Lyapunov exponent for simplicity. We calculated $\lambda_{c}$ and plotted it as a function of the drive strength $D$ in Figs. 1(c) and 1(d). It can be seen that the onset of consistency ($C=1$) corresponds to a negative value of the maximum conditional Lyapunov exponent, which indicates that close trajectories tend to converge on the average when the system is driven strongly enough.

For comparison, we also calculated the usual Lyapunov exponent $\lambda$ by using the linearized equations of the whole drive-response system for the chaos-driven Lorenz model; $d{\mbox{\boldmath $\eta$}}(t)/dt = {\mbox{\boldmath D}}_{\tiny \mbox{\boldmath X,S}}{\mbox{\boldmath F}}({\mbox{\boldmath $\eta$}}(t), {\mbox{\boldmath X}}(t), {\mbox{\boldmath S}}(t))$, where ${\mbox{\boldmath $\eta$}}$ are the linearized variables of both ${\mbox{\boldmath X}}$ and ${\mbox{\boldmath S}}$, and ${\mbox{\boldmath D}}_{\tiny \mbox{\boldmath X,S}}{\mbox{\boldmath F}}$ is the Jacobian matrix with respect to both ${\mbox{\boldmath X}}$ and ${\mbox{\boldmath S}}$. $\lambda$ cannot be calculated for noise-driven systems because of infinite dimensionality of the noise drive signal $S(t)$. $\lambda$ is plotted as a function of the drive strength $D$ as shown in Fig. 1(c). It is worth noting that the measure of consistency $\lambda_{c}$ is different from the measure of chaoticity $\lambda$ in the presence of drive signals. At large drive strengths consistent behavior can be observed ($\lambda_{c} < 0$) even for a chaotic motion ($\lambda > 0$).

In order to see the convergence behavior locally in time and space, we introduce the local conditional Lyapunov exponent (LCLE),
\begin{eqnarray}
\label{equation7}
\lambda_{lc}(t)=\frac{1}{\Delta t}\log\frac{\|{\mbox{\boldmath $\xi$}}(t+\Delta t)\|}{\|{\mbox{\boldmath $\xi$}}(t)\|},
\end{eqnarray}

where $\Delta t$ is a finite time corresponding to the time step of the numerical calculation. ${\mbox{\boldmath $\xi$}}(t)$ is calculated along the trajectory of the response system, which is influenced by the time-dependent drive signal. $\|{\mbox{\boldmath $\xi$}}(t)\|$ is normalized at each step of numerical integration so that the vector ${\mbox{\boldmath $\xi$}}(t)$ stays a unit vector. It is important to maintain the direction of the vector ${\mbox{\boldmath $\xi$}}(t)$ to incorporate the influence of the time-dependent trajectory in the phase space.

We plotted $\lambda_{lc}$ on the trajectory of the Lorenz `butterfly' attractor on the $x$-$z$ plane of the phase space. Figure 2(a) shows $\lambda_{lc}$ before a drive signal is added. Figures 2(b) and 2(c) show $\lambda_{lc}$ for a response driven by a Lorenz chaos and a colored noise drive signal, respectively. Strong expansion ($\lambda_{lc}>0$) and contraction ($\lambda_{lc}<0$) regions, are indicated by red and blue regions. The red regions show the locations on the attractor where the system is more easily susceptible to breakdown of consistency due to perturbative noise. Comparing the Figures 2(a), (b) and (c) it can be seen how the distribution of converging and diverging dynamics can depend on the drive signal. Zhou and Kurths discussed contraction regions in the driven butterfly in the context of synchronization~\cite{zhou230602} using a definition of contraction region based on he Jacobian matrix of the response system, rather than the LCLE. This approach does not incorporate the dependence of the convergence on the time dependent nature of the drive signal.

We show the change in shape of the distribution of $\lambda_{lc}$ values for increasing drive strength $D$ in Fig. 3(a). The distributions of $\lambda_{lc}$ at constant drive strengths $D=0$, 60 and 120 are shown in Figs. 3(b)-3(d). These figures correspond to the average values of $\lambda_{c}$ = 0.905, -0.110 and -1.507, respectively. The shape of the distribution shifts towards negative values of $\lambda_{lc}$ with increase of drive strength. However, the position of the main peak, corresponding to weak contractions, does not change, and the onset of consistency does not require a drastic shift of the whole distribution to the negative side.


As a final point of discussion, we note that the LCLE measures the convergence of close trajectories, and does not guarantee the convergence of trajectories from arbitrary positions in the phase space. A response system could have multiple basins - that could be more than one response attractor. An example is the case of $r=13$, where the Lorenz system has two fixed points without a drive signal. When we drive the $x$ variable with a colored-noise signal of small drive strength, $D=10$, (and $\tau_{c}=2.0$) we can obtain two types of response trajectories for the same drive signal depending on the initial conditions. Figure 4(a) shows two separate response trajectories (black curves) located near the original fixed points on the $x$-$y$ plane of the phase space. The color code shows the probability of converging into the consistent trajectory in one of the two basins for a set of colored-noise drive signals (ten different temporal waveforms). The basins of convergence for each of the two possible responses are shown in red and blue. Each of these responses can be seen to be locally consistent in the sense that trajectories starting in the same basin converge after transients. The boundary of the two basins depends on the drive signal. This example shows that the ensemble of initial states is significant for the evaluation of consistency measures as defined in Eq. (1). When the drive strength is increased to $D=30$ the amplitudes of the two attracting response trajectories become larger and merge together. For large drive strengths there exists only one attracting response trajectory and a single basin of consistency for all initial conditions, as shown in Fig. 4(b).

In conclusion, we have introduced the viewpoint of consistency in driven nonlinear dynamical systems and used the Lorenz model driven by chaos and colored noise signals to illustrate this viewpoint. Consistency is evaluated by the cross-correlation function between responses starting from different initial conditions when the system is repeatedly driven by the same waveform. Consistent responses are observed after transients even though the system is in a chaotic state. A rather different insight is provided when we compute and plot the LCLE in the phase space, which allows us to visualize contraction and expansion regions on the driven chaotic attractors. We observe the dependence of the distribution of the LCLE on changes of the drive signal to understand the onset of consistency. We also found the multiple basins of consistency, where the Lorenz system exhibits more than one type of response with respect to the same drive signal, and yet is still consistent in the local sense that the system converges to the same response for large sets of initial states. We expect that the viewpoint of consistency in driven nonlinear dynamical systems will be useful for analyzing and understanding drive-response characteristics of complex systems in science and engineering.

We thank Lou Pecora and Arkady Pikovsky for helpful comments and criticism. We thank Masato Miyoshi, Shoji Makino, and Naonori Ueda for their support and encouragement. A.U. gratefully acknowledges support from NTT Corporation, the Support Center for Advanced Telecommunications Technology Research, and Grants-in-Aid for Scientific Research from the Japan Society for the Promotion of Science. R.R. gratefully acknowledges travel support from NTT Corporation for this collaboration. 



\begin{figure}[h]
\caption{(a),(b) Temporal waveforms of the drive signal and the x variable of the driven Lorenz model starting from ten different initial conditions. (a) Chaos drive and (b) colored noise drive. (c),(d) The consistency $C$ (black curve), the conditional Lyapunov exponent $\lambda_{c}$ (blue curve),  and the Lyapunov exponent $\lambda$ (red line) as a function of drive strength $D$. (c) Chaos drive and (d) colored noise drive. $\sigma=10$, $r=28$, and $b=8/3$ are often used for the Lorenz model to observe chaotic behaviors without a drive signal. $y$ variable of the Lorenz model is additively driven by a Lorenz chaos and a colored noise signal with a correlation time $\tau_{c}=0.2$.}
\label{fig1}
\end{figure}

\begin{figure}[h]
\caption{The driven butterfly. (a) No drive, (b) chaos drive ($D=15$), and (c) colored noise drive ($D=70$). The color indicates the local conditional Lyapunov exponent plotted on the trajectory in the x-z plane of the phase space. Expansion (red) and contraction (blue) regions are observed on the trajectory.}
\label{fig2}
\end{figure}

\begin{figure}[h]
\caption{(a) Three dimensional picture of the distribution of local conditional Lyapunov exponent $\lambda_{lc}$ as a function of colored-noise drive strength $D$. (b)-(d) Distribution of $\lambda_{lc}$ at constant drive strengths. (b) $D=0$, (c) $D=60$, and (d) $D=120$. The conditional Lyapunov exponent $\lambda_{c}$ is obtained from the average of the distribution of $\lambda_{lc}$: (b) $\lambda_{c} = 0.905$, (c) $\lambda_{c} = -0.110$, and (d) $\lambda_{c} = -1.507$.}
\label{fig3}
\end{figure}

\begin{figure}[h]
\caption{(a) Multiple basins of consistency at D = 10. Black lines show attracting trajectories which correspond to locally consistent responses. Color indicates the probability of converging into a trajectory in the basin 1, plotted on the $x-y$ plane of the three-dimensional phase space at $z=12$. Red corresponds to the basin 1 and blue corresponds to the basin 2. (b) Single basin of consistency at D = 30. Black line indicates the consistent trajectory. Red indicates the basin 1. The Lorenz model has two fixed points without a drive signal ($r$ = 13). $x$-variable of the Lorenz model is additively driven by a colored-noise signal with a correlation time $\tau_{c}=2.0$.}
\label{fig4}
\end{figure}

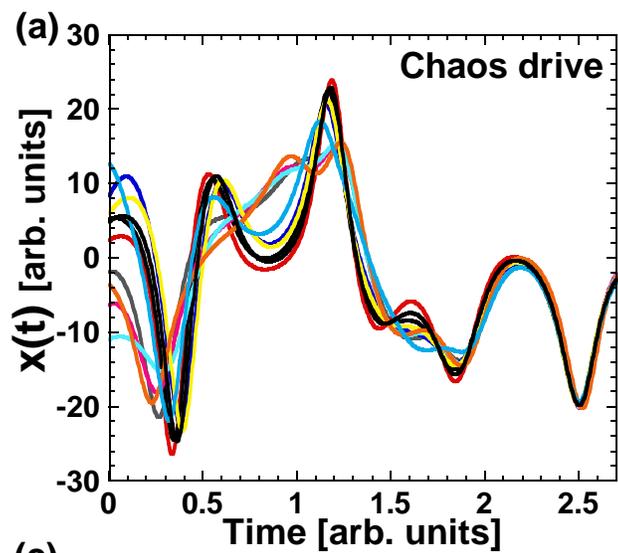
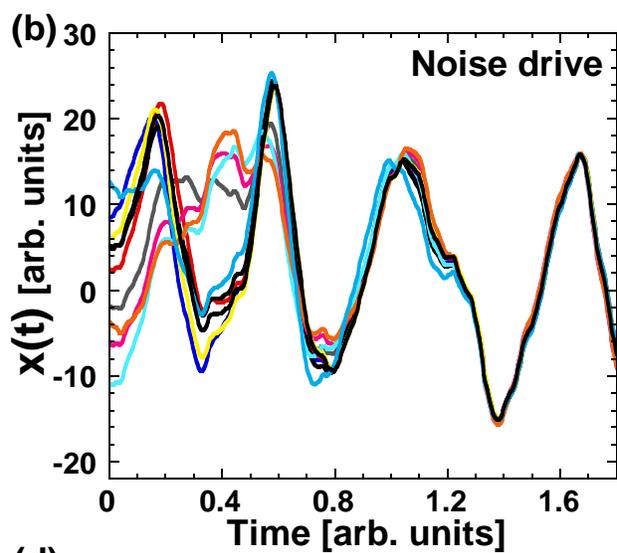
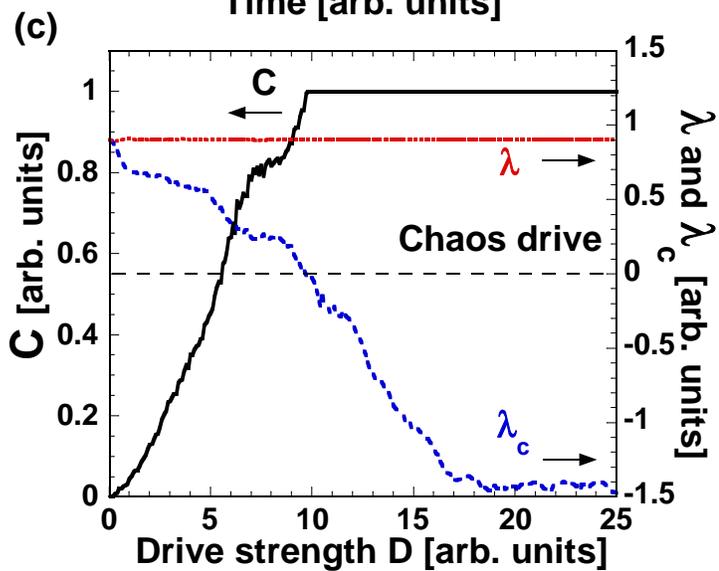
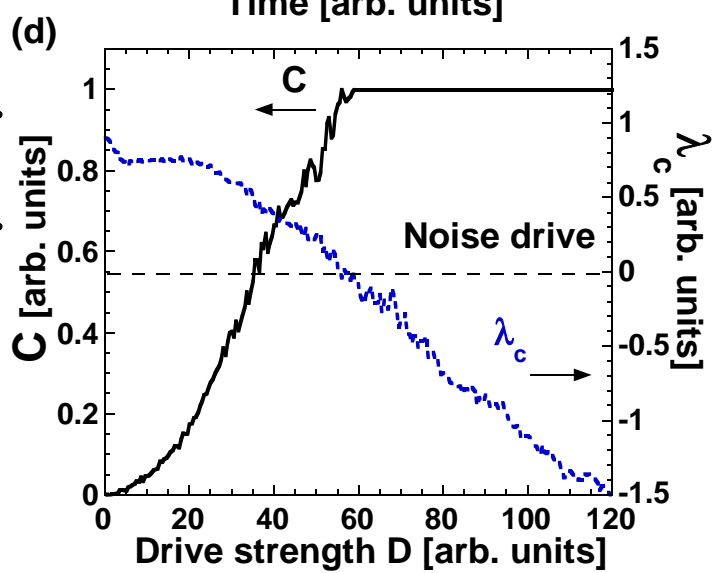
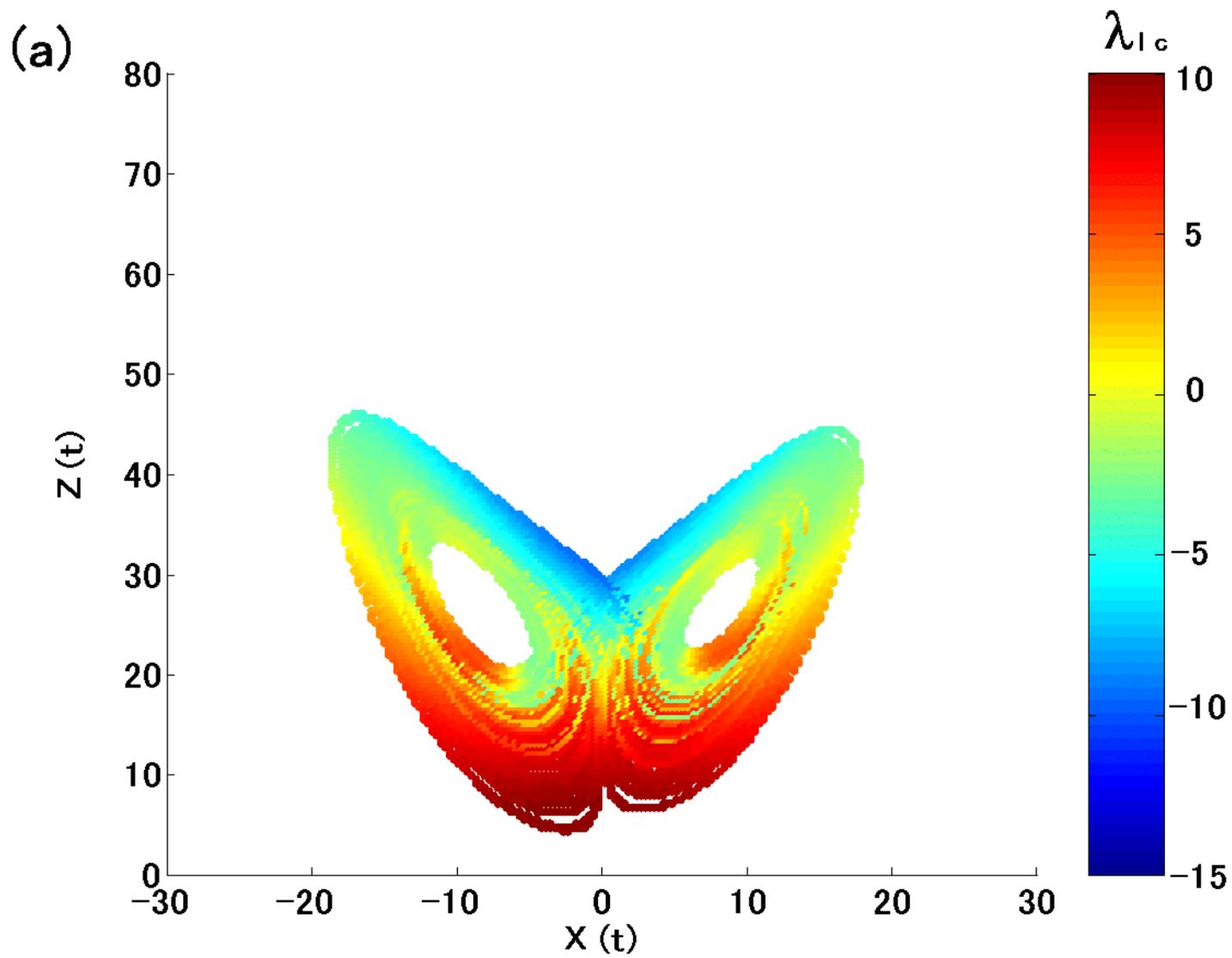
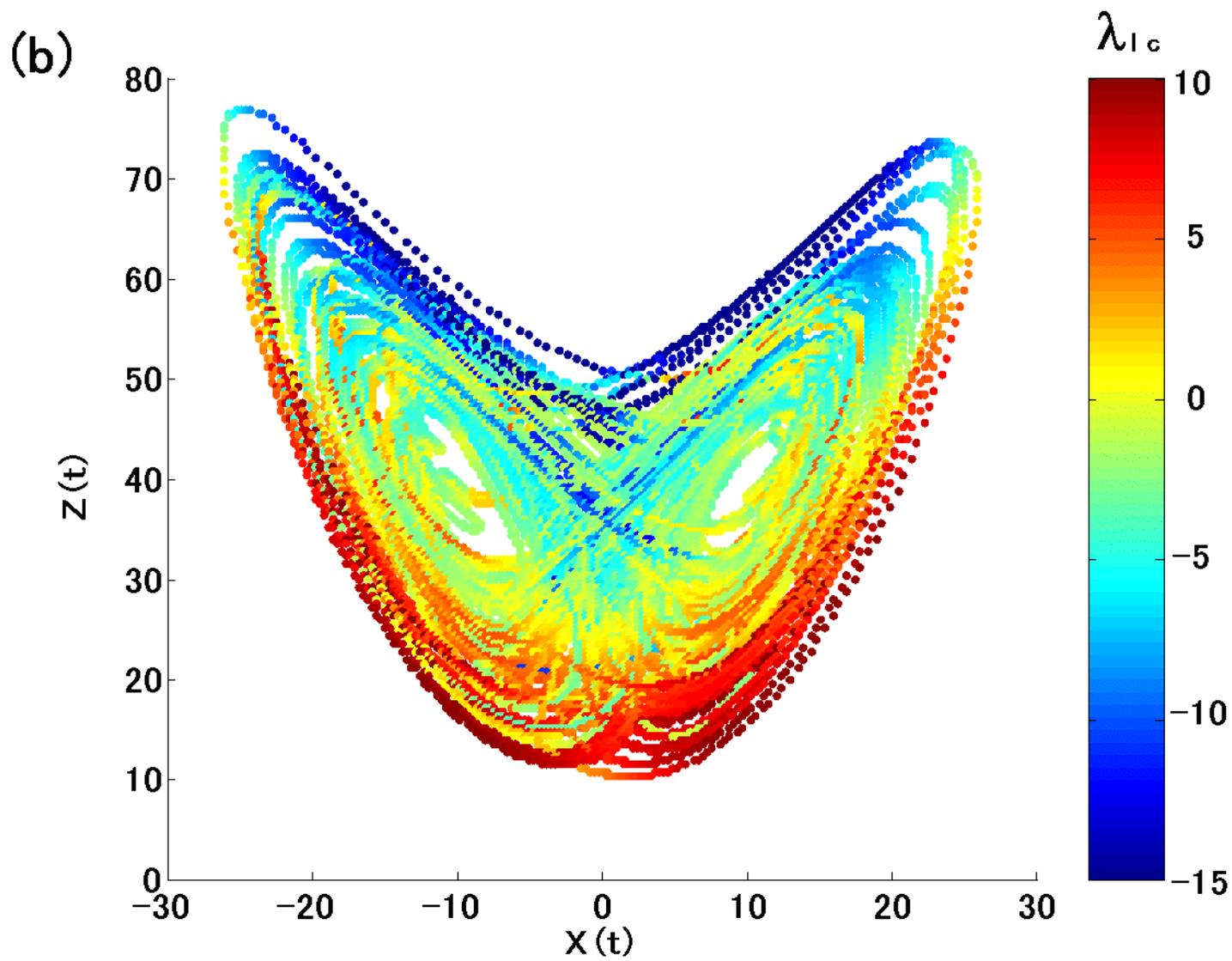
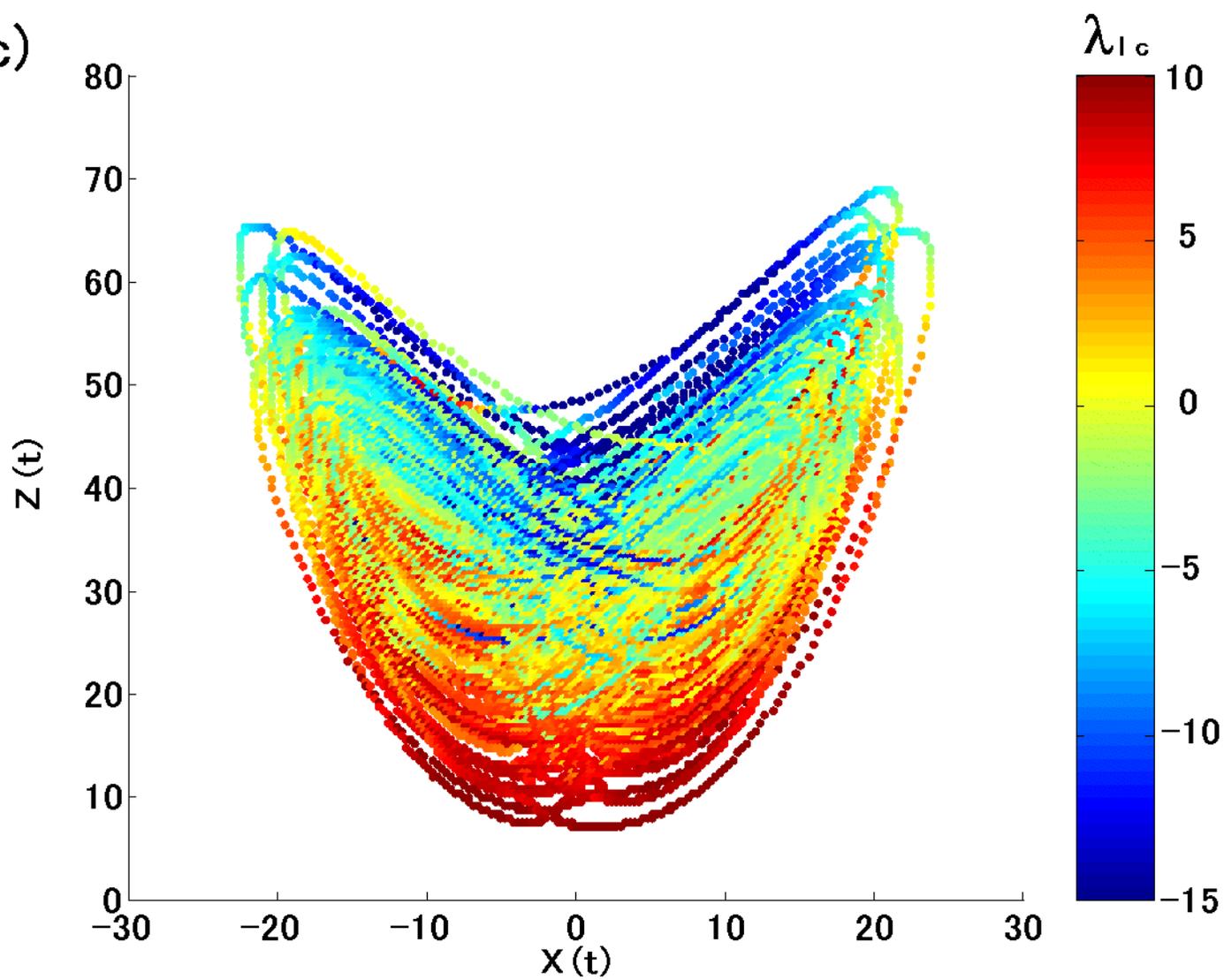
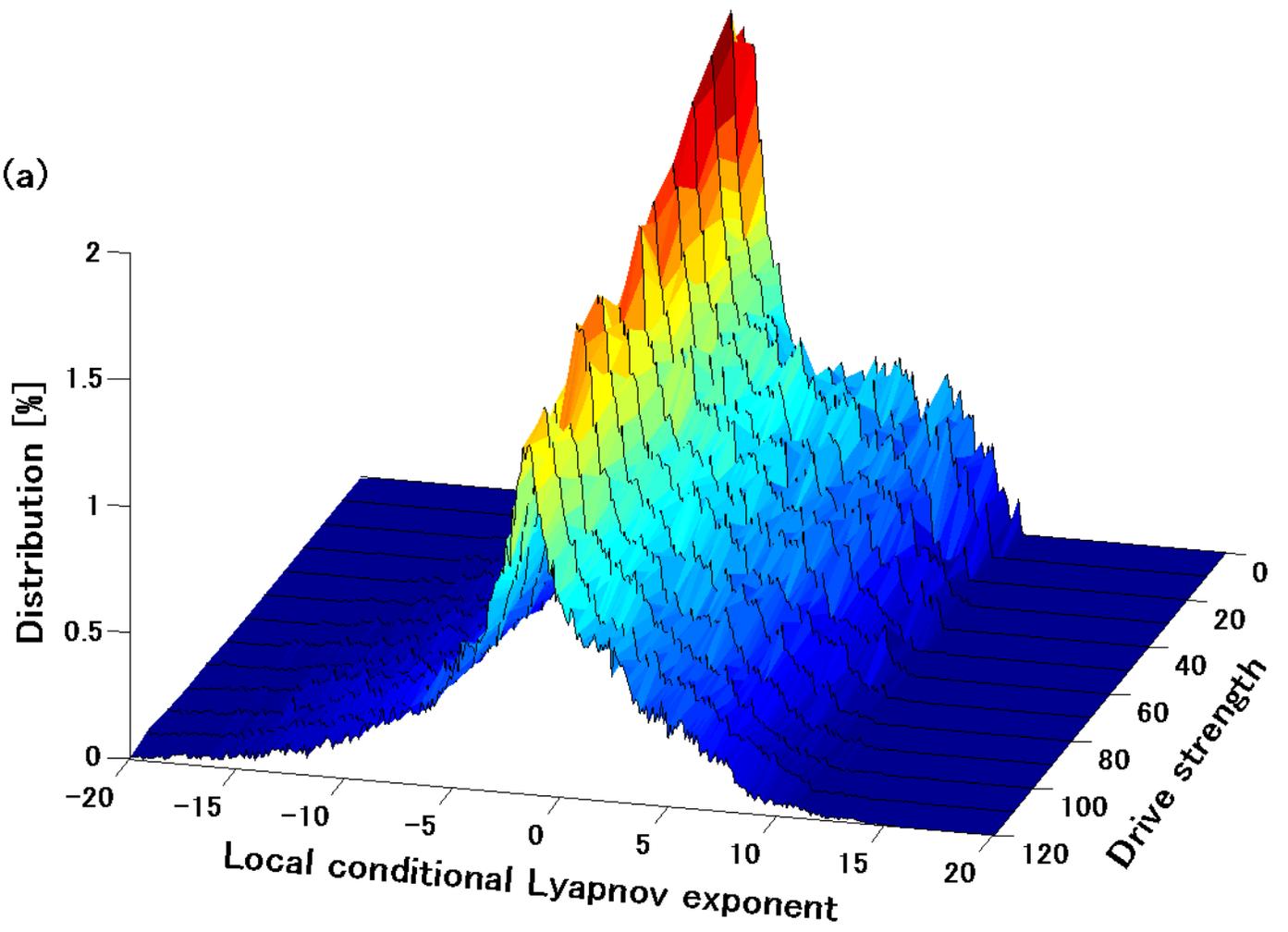
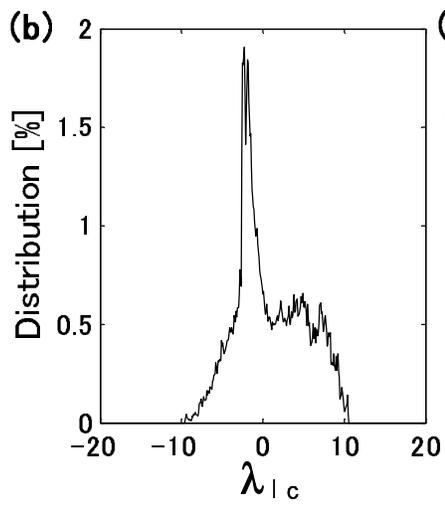
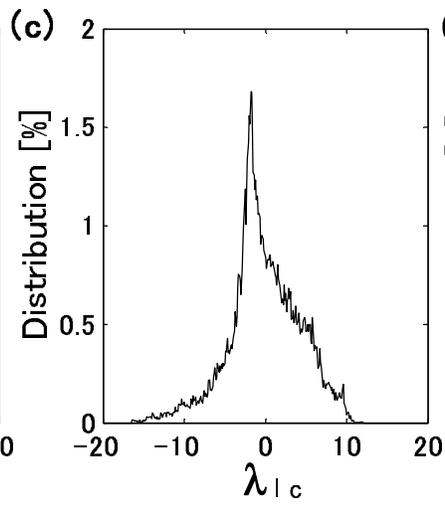
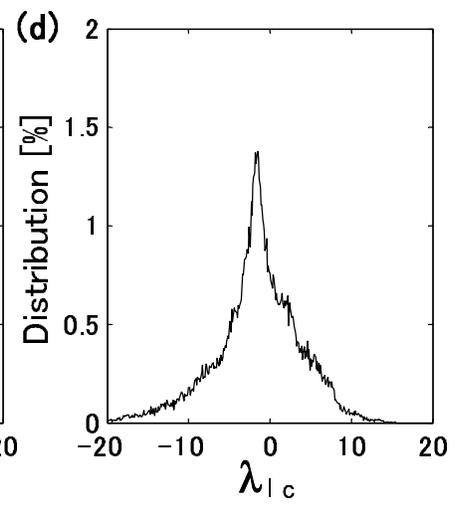
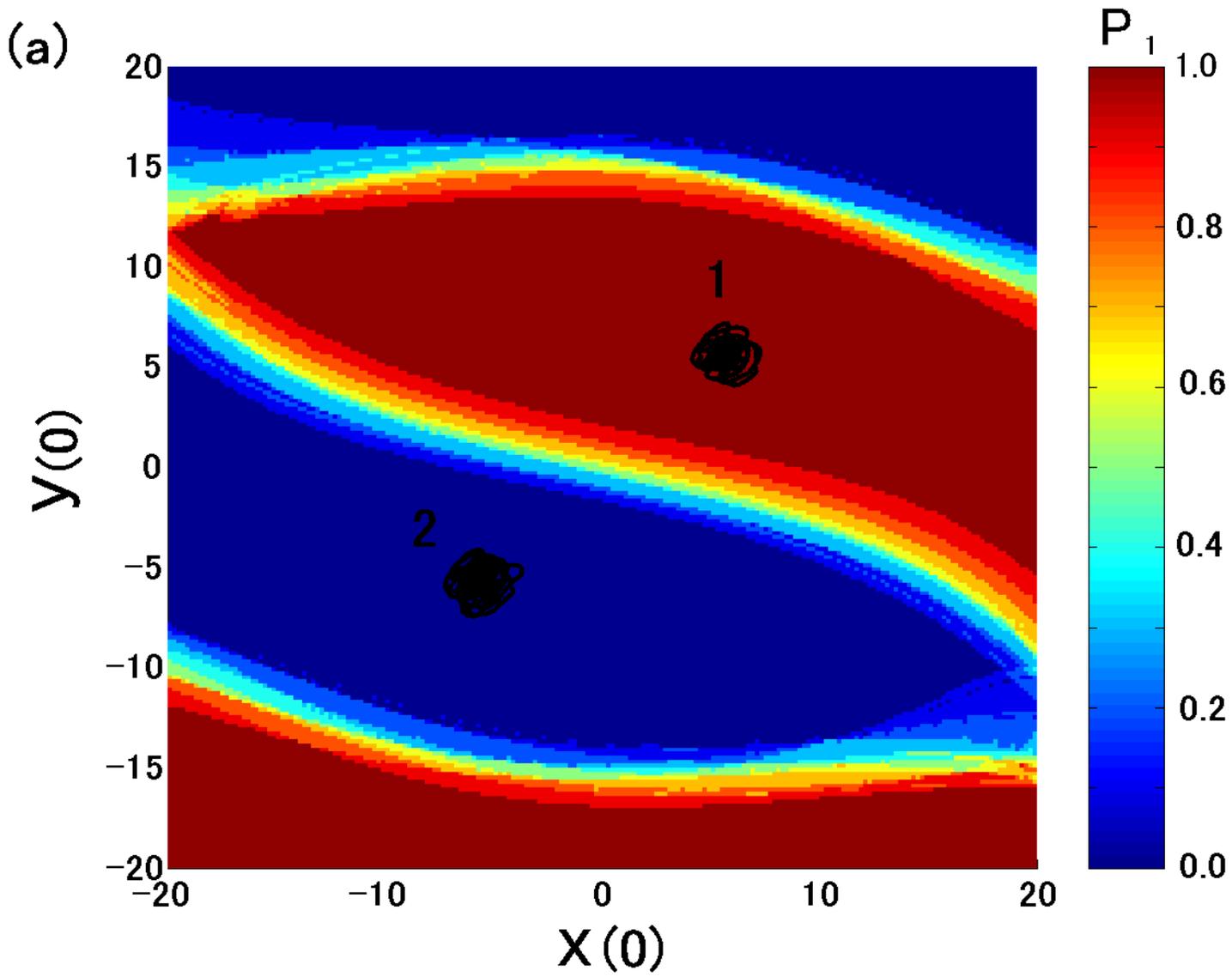
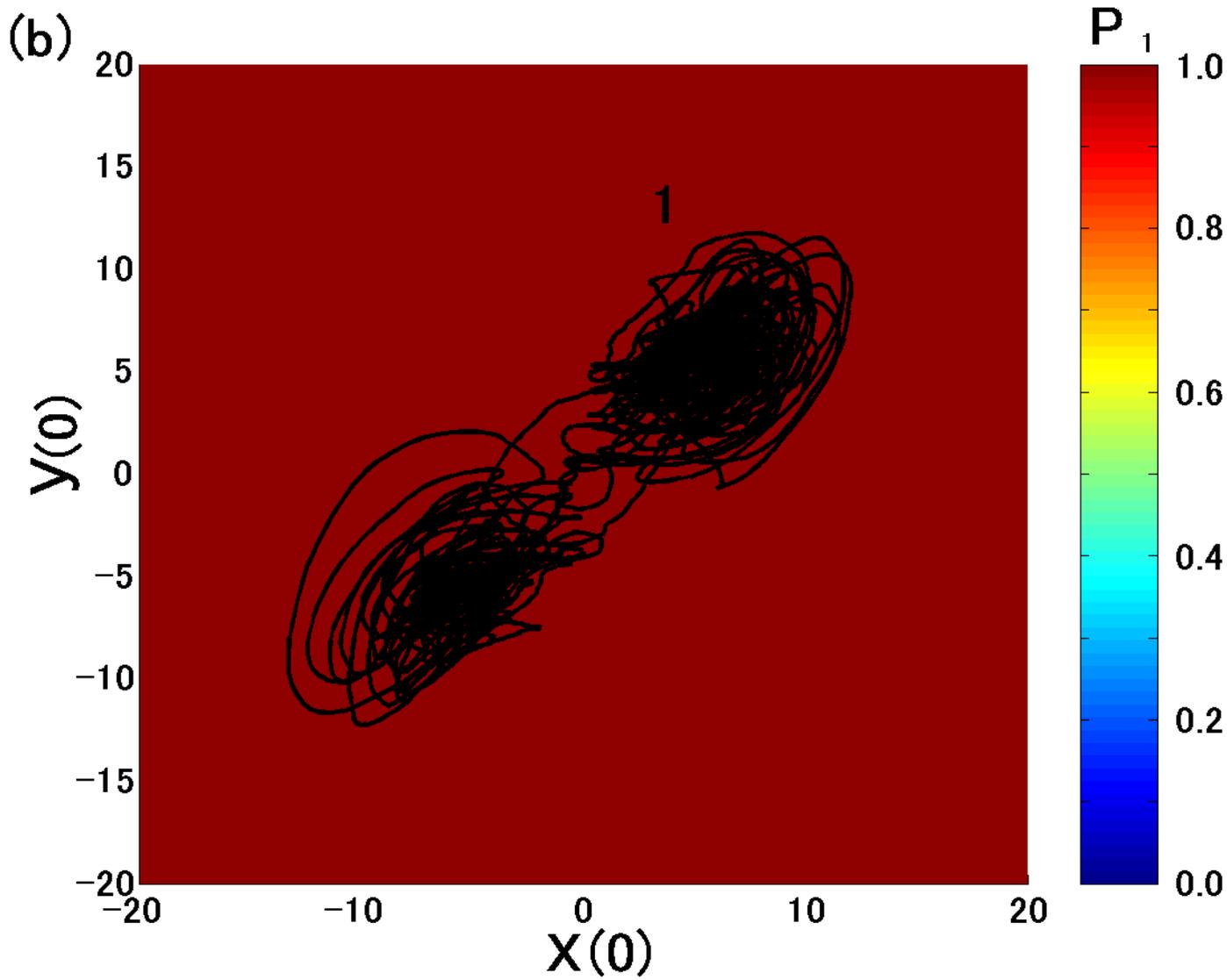
\end{document}